\documentclass{appolb}
\usepackage{graphicx}
\usepackage{amsmath}
\usepackage{cite}
\headtitle{Quantum Brownian motion and the Third Law of thermodynamics}
\headauthor{Peter H{\"a}nggi and Gert-Ludwig Ingold}
\begin{document}
\title{Quantum Brownian motion and the Third Law of
thermodynamics
\thanks{This work is
dedicated to our colleague and  friend, Professor Peter Talkner, on
the occasion of his 60-th birthday.}}
\author{Peter H{\"a}nggi and Gert-Ludwig Ingold
\address{Institut f{\"u}r Physik, Universit{\"a}t Augsburg, 86135 Augsburg,
Germany}
}
\maketitle
\begin{abstract}
The quantum thermodynamic behavior of small systems is investigated
in presence of finite quantum dissipation. We consider the archetype
cases of a damped harmonic oscillator and a free quantum Brownian
particle. A main finding is that  quantum dissipation helps to ensure the validity of
the Third Law. For the quantum oscillator, finite damping replaces the
zero-coupling result of an exponential
suppression of the specific heat at low temperatures by a
power-law behavior. Rather intriguing is the behavior of the free quantum
Brownian particle. In this case, quantum dissipation is able to restore the Third Law:
Instead of being constant down to zero temperature, the specific heat now
vanishes proportional to temperature with an amplitude that is {\it inversely}
proportional to the ohmic dissipation strength. A distinct subtlety of
finite quantum dissipation is the result that the various thermodynamic functions of
the sub-system do not only depend on the dissipation strength but depend as well
on the prescription employed in their definition.
\end{abstract}

\section{Introduction}
The development of the theory of Brownian motion played  a pivotal
role --  and continues to do so -- in the development of statistical
mechanics and thermodynamics \cite{chaos05,physicaSI}.  Thermodynamics
together with relativity and quantum theory form three pillars on
which much of the entire structure of physics rests.
Tampering with the axioms in either of those theories is not a good
idea; doing so may well lead to contradictions with
the  other theories. In particular, the field of thermodynamics
bears consequences for many branches of physics. Its four laws are
well-known \cite{TD1,TD2}: the zeroth law guarantees that states of
thermal equilibrium exist which can be characterized by
a temperature $T$. The first law provides a balance among the
various contributions that make up the internal energy of a system
while the second law introduces the concept of  thermodynamic entropy $S$,
which notably is extensive and never decreases for a closed physical
system. In addition, the second law tells us that there exists an
absolute zero of temperature.

The Third Law is attributed to Walther Hermann Nernst (1864-1941)
and arose as the result of his seminal idea -- being guided by his
critical analysis of  chemical and electrochemical reactions at
lower temperatures -- that at low temperatures there occurs for
isothermal processes a perfect correspondence between the enthalpy
and the Gibbs free energy. Thereby, the approximate rule
hypothesized earlier by Marcelin Berthelot and Julius Thomson, becomes
a ``law'' at zero
temperature. Nernst announced this result already in his lectures in
1905, terming it ``mein W\"armesatz'' (my law of heat)
\cite{ebeling1,ebeling2}. He took this result even further: He also
studied {\it how} fast the difference between the changes in the
enthalpy $\Delta H$ and the Gibbs free energy $\Delta G$, \ie
$\Delta H - \Delta G$ tends to zero \cite{Nernst}.  In fact, this
difference vanishes faster than linear in temperature implying that
the change of entropy itself must vanish at absolute zero. This in
turn implies identical, generally vanishing initial slopes for the corresponding
quantities $\Delta H$ and $\Delta G$ as a function of temperature.
An elucidating account of the history of the Third Law and the
controversies surrounding its acceptance is presented in the books
by Dugdale \cite{Dugdale} and by Wilks \cite{Wilks}. In its strict form,
as given by Max Planck
\cite{planck}, the Third Law reads: The entropy $s=S/N$ per particle
approaches at absolute zero a constant value $s_0$ that possibly depends
on the chemical composition of the system. This limiting entropy constant
$s_0$ can generally be set equal to zero.

The Third Law carries prominent consequences for quantum mechanics
and the field of low-temperature physics. First, the fact
that at absolute zero temperature the isotherm coincides with the isentrope (adiabat)
immediately implies that this absolute zero temperature is
unattainable by  use of a sequence of  isothermal and
adiabatic reversible operations \cite{nernst1912}.
Therefore, it has the consequence that
the efficiency of a Carnot engine, \ie a heat engine that cyclically
operates between two heat baths of different temperatures which are
never brought into contact with each other, can never reach $100\%$ for
any finite upper temperature. Moreover, the constant
value of the entropy at absolute zero is given by the degeneracy $g$ of the $N$-particle system
in the corresponding quantum ground state, \ie $S(T=0)= k_\text{B}\ln g$,
where $k_\text{B}$ is the Boltzmann constant. The limiting value of the
intensive quantity $s_0=S(T=0)/N$ in the thermodynamic limit of
particle number $N \rightarrow \infty$ will typically be zero,
so long as the degeneracy $g=g(N)$ does not grow with $N$ faster than
exponentially \cite{Leggett}. A well-known exception is the case of
noninteracting, independent particles carrying a non-vanishing spin $I$, yielding
$s_0=k_\text{B}\ln(2I+1)$ for the limiting entropy per particle.
Moreover, the Third Law also implies
that thermal quantities such as specific heats, the isobar thermal
coefficient of expansion, the isochor coefficient of tension, etc.,
all approach zero as $T \rightarrow 0$. Likewise, the (magnetic)
susceptibility becomes constant as $T \rightarrow 0$, so that
the classic Curie law must loose its validity at very small
temperatures.

Are there known exceptions of systems not obeying the Third Law? It
is known that many classical systems do not obey the Third Law. In
particular, noninteracting classical systems with their constant
values for the specific heat clearly violate the Third Law. A
well-known case is the classical ideal gas for which the entropy $S$
assumes the form $S=N[c_V\ln T+k_\text{B}\ln(V/N)+\sigma]$, where
$V$ is the volume and
$\sigma$ denotes the entropy constant. It clearly does not fulfill
the Third Law because it diverges logarithmically with temperature $T$
for a constant specific heat $c_V$. Even when we use for $c_V$ the
physically relevant low-temperature result, namely that quantum
mechanically the specific heat $c_V$ vanishes faster than $\ln T$,
we still find a dependence on the particle density which is not compatible 
with the formulation of Planck. This observation that the classical gas
does not obey the Third Law led Nernst to speculate that the
classical gases must undergo a ``degeneracy'', which has been
resolved with the quantum statistics for the ideal Fermi gas and the
ideal Bose-Einstein gas, which indeed do obey the Third Law
in the strict formulation by Planck.

Are there yet other remaining open problems with the Third Law?
Apparent difficulties with the Third Law occur for metastable states
that do not necessarily guarantee the sufficiently fast relaxation
within a finite time scale towards thermal equilibrium, the latter
being a prerequisite for the validity of the Third Law. In this
context, glasses provide a system class that can provide
detectable deviations from the Third Law at low temperatures which
likely are the result of frozen-in ordered excited states that have
not yet fully relaxed. According to common wisdom the known
deviations from the Third Law will all be cured by quantum mechanics,
quantum statistics, and interactions among particles.

In the following we shall investigate the thermodynamic low
temperature properties for open quantum systems that are coupled to
a heat bath of {\it finite} dissipation strength. Because both, the
thermodynamics of classical open systems and the quantum statistical
mechanics of open systems are strictly valid only for systems that
are only infinitely weakly coupled to a bath, it is
{\it a priori} not obvious how the Third Law lives up to the
sub-system in presence of finite quantum dissipation
\cite{leggett83,rise85,ingoldPR,HTB,schon90,dittrich98,grifoni98,weiss99,talkner04,chaosHI}.
The effect of the finite coupling of a sub-system to an environment
in fact induces several subtleties for quantum Brownian motion
\cite{chaosHI}. For example, the equilibrium density matrix
is no longer given by its standard canonical form $\rho_\text{S}=
\exp (-\beta H_\text{S})/Z$ where $H_\text{S}$ denotes the
system Hamiltonian and $Z=\Tr[\exp(-\beta H_\text{S})]$
is the partition function. Therefore, in clear contrast to the classical case
with dissipation, this reduced density matrix becomes also a
function of the interaction strength with the environment. This
being so, taking a gas of free independent quantum Brownian
particles that are coupled to a heat bath with finite dissipation
strength, one may speculate that the role of the interactions
of the free particle with the abundant bath degrees of freedom will
be sufficient to cure the shortcomings stemming from a classical gas
of free Brownian particles.

\section{Quantum harmonic oscillator revisited}
To start out, we first recall the standard results for a single
harmonic quantum oscillator that is infinitely weakly coupled
to a bath that establishes the temperature $T$. This
situation is reminiscent of the famous treatment of the specific
heat of a solid by Albert Einstein in 1907 \cite{einstein07},
where he found an exponential suppression of the specific heat
as $T \rightarrow 0$. This finding impressed the Berlin
school so immensely, in particular Nernst and his collaborators
(who in 1910 experimentally confirmed this salient first
prediction of quantum theory), that Nernst together with Planck were
able to bring the ``new Copernicus'' \cite{ebeling1,ebeling2}
into the exclusive circle of Berlin physicists in 1913.

\subsection{Partition function and entropy}
\label{pfe}
Let us consider an oscillator degree of freedom of mass $M$ and
force constant $f$, \ie its Hamiltonian $H_\text{S}$ reads
\begin{equation} \label{HO}
  H_\text{S} = \frac{p^2}{2M} + \frac{1}{2} fx^2\,.
\end{equation}
In terms of the angular frequency $\omega_0^2= f/M$ the quantum mechanical
energy eigenvalues read $E_n= (n+ \frac{1}{2}) \hbar \omega_0$, yielding
for the partition function $Z$ the well-known expression
\begin{equation} \label{PHO}
  Z= \sum_{n=0}^\infty \e^{-\beta E_n} =
  \frac{1}{2 \sinh[\hbar\beta\omega_0/2]}
\end{equation}
where $\beta=1/k_\text{B}T$ is the inverse temperature.

Using familiar relations we find that the internal energy  $E$ reads
\begin{align} \label{HOE}
  E &= -\frac{\partial}{\partial\beta}\ln(Z)\\
    &= \frac{\hbar\omega_0}{2} +\frac{\hbar\omega_0}{\exp(\hbar\beta\omega_0)-1}
\end{align}
and, correspondingly, the entropy is given by
\begin{align} \label{HOS}
  S &= k_\text{B}\left[\ln(Z)-\beta\frac{\partial}{\partial\beta}\ln(Z)\right]\\
  &= k_\text{B}\left[\frac{\hbar\beta\omega_0}{\exp(\hbar\beta\omega_0)-1}-
       \ln\big(1-\exp(-\hbar\beta\omega_0)\big)\right]\,.
  \label{HOSe}
\end{align}
For low temperatures, $\hbar\beta\omega_0\gg1$, the entropy
approaches zero like
\begin{equation}
  S = \frac{\hbar\omega_0}{T}\exp\left(-\frac{\hbar\omega_0}{k_\text{B}T}\right)\,.
\end{equation}

The specific heat can now be derived either from (\ref{HOE}) as
\begin{equation}\label{CE}
  C=\frac{\partial E}{\partial T}=-k_\text{B}\beta^2\frac{\partial E}{\partial\beta}
\end{equation}
or from (\ref{HOS}) as
\begin{equation}\label{CS}
  C=T\frac{\partial S}{\partial T}=-\beta\frac{\partial S}{\partial\beta}\,.
\end{equation}
In both cases, one obtains for the specific heat
\begin{equation}
  C=k_\text{B}\left(\frac{\hbar\beta\omega_0}{2\sinh(\hbar\beta\omega_0/2)}\right)^2\,.
\end{equation}
Its low-temperature behavior
\begin{equation}\label{CLT}
  C=  k_\text{B} \left(\frac{\hbar\omega_0}{k_\text{B}T} \right)^2
  \exp\left(-\frac{\hbar\omega_0}{k_\text{B}T}\right)
  \,,
\end{equation}
is not analytic in temperature and corresponds to Einstein's result
for the low-temperature behavior of the specific heat of a solid
\cite{einstein07}. For high temperatures one finds
\begin{equation}
  C = k_\text{B}\left[1-\frac{1}{12}\left(\frac{\hbar\omega_0}
  {k_\text{B}T}\right)^2+O(T^{-4})\right]\,.
\end{equation}
As this result shows, the specific heat for a free particle cannot simply
be obtained by taking the limit $\omega_0\to0$ of the harmonic oscillator.
Such a procedure will not properly account for the reduced number of degrees of
freedom which within the equipartition theorem will lead to a high-temperature 
specific heat of only $C=k_\text{B}/2$ for the free particle.

\section{Quantum harmonic oscillator: The role of quantum dissipation}
\subsection{Harmonic oscillator coupled to an environment}

We now couple the harmonic oscillator of the previous section to an
environment consisting
of an infinite number of harmonic oscillators forming a heat bath. In
contrast to the previous section, the coupling strength will not be
kept negligible here. In addition, system and bath are infinitely
weakly coupled to a superbath which has the purpose to provide
the temperature $T$.

The total Hamiltonian $H$ does not need to account for the superbath
and therefore consists of three parts
\cite{ingoldPR,HTB,dittrich98,chaosHI,mag,connell}
\begin{equation} \label{total}
  H = H_\text{S}+H_\text{B}+H_\text{SB}
\end{equation}
where $H_\text{S}$ is given by (\ref{HO}), the bath Hamiltonian reads
\begin{equation}
  H_\text{B} =
  \sum_{i=1}^\infty\left(\frac{p_i^2}{2m_i}+\frac{m_i\omega_i^2}{2}x_i^2\right)
\end{equation}
and the coupling is bilinear in the coordinates
\begin{equation}
  H_\text{SB}= -q\sum_{i=1}^{\infty}c_ix_i + q^2\sum\frac{c_i^2}{2m_i\omega_i^2}\,.
\end{equation}

We note that quantum systems that are coupled to an environment of finite strength
are rarely exactly solvable. The dissipative quantum oscillator  becomes exactly solvable
with its bilinear coupling to a bath because of the inherent quadratic structure of the total
Hamiltonian in (\ref{total}). This fact holds true even for the case of time-dependent,
parametrically driven dissipative quantum harmonic systems \cite{zerbe,kohler}.

In order to describe the influence of the environment on the system
oscillator, it is sufficient to know the spectral density of bath
oscillators defined by \cite{leggett83,ingoldPR,HTB,grifoni98}
\begin{equation}
  \label{jom}
  J(\omega) = \pi\sum_{i=1}^{\infty}\frac{c_i^2}{2m_i\omega_i}
     \delta(\omega-\omega_i)\,.
\end{equation}
For later purposes, we introduce the Laplace transform of the
damping kernel, which generally depends on frequency, thereby
causing memory-friction \cite{talkner80,ingold85}, \ie,
\begin{equation}
  \hat\gamma(z) = \frac{1}{M}\int_0^{\infty}\frac{\text{d}\omega}{\pi}
  \frac{J(\omega)}{\omega}\frac{2z}{\omega^2+z^2}\,.
\end{equation}
The important special case of strictly ohmic dissipation is characterized by
$J(\omega)=M\gamma\omega$ and $\hat\gamma(z)=\gamma$ which leads to a
memoryless damping of strength $\gamma$.

\subsection{Specific heat of a damped harmonic oscillator}
\label{cdho}
We next discuss the specific heat of the damped harmonic oscillator
by following two routes. First, we  start from the energy $E$ and
employ the common relation in (\ref{CE}). As an alternative route
we shall in Section~\ref{evspfdho} determine the
entropy from the partition function by means of (\ref{HOS}) from which (\ref{CS})
allows one to evaluate the specific heat in the case of strictly ohmic damping.

The energy of the damped harmonic oscillator is given by
\begin{equation}
  \label{EHO}
  \langle E\rangle = \frac{\langle p^2\rangle}{2M}+\frac{M}{2}\omega_0^2
  \langle q^2\rangle
\end{equation}
where the expectation value of an operator $O_\text{S}$ acting in the Hilbert
space of the system is defined with respect to the canonical density matrix of
system plus environment as
\begin{equation}
  \langle O_\text{S}\rangle = \frac{\Tr\left[O_\text{S}\exp(-\beta H)\right]}
  {\Tr\left[\exp(-\beta H)\right]}\,.
\end{equation}
For ohmic damping, the second moments of position and momentum can
be expressed as \cite{ingoldPR,talkner84,riseborough85}
\begin{equation}
  \langle q^2\rangle=\frac{\hbar}{M}f_0(T)
\end{equation}
and
\begin{equation}
  \langle p^2\rangle=M\hbar f_2(T)
\end{equation}
where we have introduced a temperature-dependent function
\begin{equation}
  \label{fnT}
  f_n(T)=\int_{-\infty}^{+\infty}\frac{\text{d}\omega}{2\pi}
  \frac{\gamma\omega^{n+1}}{(\omega^2-\omega_0^2)^2+\gamma^2\omega^2}
  \coth\left(\frac{\hbar\beta\omega}{2}\right)\,.
\end{equation}
For $n=2$, \ie when evaluating $\langle p^2\rangle$, the integrand
decreases only with $1/\omega$ and a finite value can only be obtained
by introducing a high-frequency cutoff in the damping kernel
$\hat\gamma(z)$. However, this divergent term gives rise only to a
temperature-independent contribution to $\langle p^2\rangle$ and thus to
the energy (\ref{EHO}). When evaluating the specific heat according to
(\ref{CE}), this constant term will disappear and a finite result is
obtained even for ohmic damping. After some algebra, one finds for the
specific heat
\begin{equation}
  \label{CHO}
  \frac{C}{k_\text{B}}=1-\frac{\hbar\beta\gamma}{2\pi}+
  \lambda_+^2\psi'\left(1+\lambda_+\right)
  +\lambda_-^2\psi'\left(1+\lambda_-\right)
\end{equation}
where
\begin{equation}
  \label{lambda}
  \lambda_\pm = \frac{\hbar\beta\omega_0}{2\pi}\left[
  \frac{\gamma}{2\omega_0}\pm\sqrt{
  \left(\frac{\gamma}{2\omega_0}\right)^2-1}\right]
\end{equation}
and $\psi'(z)$ is the trigamma function. At low temperatures, the
specific heat thus assumes the form
\begin{equation}
  \frac{C}{k_\text{B}}=
  \frac{\pi}{3}\frac{\gamma}{\omega_0}\frac{k_\text{B}T}{\hbar\omega_0}+
  \frac{4\pi^3}{15}\frac{\gamma}{\omega_0}\left[3-\left(\frac{\gamma}{\omega_0}
  \right)^2\right]\left(\frac{k_\text{B}T}{\hbar\omega_0}\right)^3+O(T^5)\,.
\end{equation}
This result differs significantly from the expression (\ref{CLT}) in
the absence of dissipation. While in the latter case, the presence
of an energy gap led to an exponential suppression of the specific
heat, we now find a linear increase with temperature. This behavior indicates
the existence of a finite density of states even at small excitation energies \cite{hanke}. Even
at high temperatures the effect of dissipation can be detected, albeit in
a less spectacular manner. The leading correction in the
high-temperature expansion reads
\begin{equation}
  \frac{C}{k_\text{B}}=1-\frac{\hbar\gamma}{2\pi k_\text{B}T}+
  \frac{\hbar^2(\gamma^2-2\omega_0^2)}{24(k_\text{B}T)^2}+O(T^{-3})
  \,.
\end{equation}
Thus, the leading correction depends on the damping
strength $\gamma$. This finding is in clear contrast to the behavior
of the quantum escape rate \cite{HTB,ingold85}: There, the leading
quantum correction to the escape rate always enhances the classical
result and is {\it independent} of the dissipation strength.

\subsection{Energy versus partition function for a damped harmonic
oscillator}
\label{evspfdho}

Another prescription to obtain the specific heat starts out from
the canonical partition function
\begin{equation}
  \label{ZHO}
  Z = \frac{\Tr\left[\exp(-\beta H)\right]}
  {\Tr_B\left[\exp(-\beta H_\text{B})\right]}\,,
\end{equation}
where $\Tr_B$ denotes the partial trace in the Hilbert space of the bath.
In the absence of a system-bath coupling, this expression would correspond
to the partition function of the system alone. For a damped harmonic
oscillator, the partition function becomes \cite{talknerFN}
\begin{equation}
  \label{ZHOe}
  Z = \frac{1}{\hbar\beta\omega_0}\prod_{n=1}^\infty \frac{\nu_n^2}
  {\nu_n^2+\nu_n\hat\gamma(\nu_n)+\omega_0^2}
\end{equation}
with the Matsubara frequencies $\nu_n=2\pi n/\hbar\beta$. In view of the
divergence for strictly ohmic damping mentioned above, we allow here for
a possible frequency dependence of $\hat\gamma$.

Following the standard procedure of statistical mechanics, we can obtain the
energy from the partition function by means of
\begin{equation}
  \label{EZZ}
  \langle E\rangle_Z = -\frac{\partial}{\partial\beta}\ln(Z)\,.
\end{equation}
Inserting (\ref{ZHOe}), one obtains
\begin{equation}
  \langle E\rangle_Z = \frac{1}{\beta}\left[1+\sum_{n=1}^\infty
  \frac{2\omega_0^2+\nu_n\hat\gamma(\nu_n)-\nu_n^2\hat\gamma'(\nu_n)}{\nu_n^2+
  \nu_n\hat\gamma(\nu_n)+\omega_0^2}\right]
\end{equation}
which in general differs from the expression
\begin{equation}
  \langle E\rangle = \frac{1}{\beta}\left[1+\sum_{n=1}^\infty
  \frac{2\omega_0^2+\nu_n\hat\gamma(\nu_n)}{\nu_n^2+
  \nu_n\hat\gamma(\nu_n)+\omega_0^2}\right]
\end{equation}
obtained by evaluating the integral (\ref{fnT}) by residues. The only
exception is the special case of strictly ohmic damping where
$\hat\gamma(\nu_n)=\gamma$ is constant.

This generally non-vanishing difference does not come as a
surprise if one only takes a closer look at
the partition function (\ref{ZHO}): The  evaluation of (\ref{EZZ}) yields
\begin{equation}
\begin{aligned}
  \label{EZvsE}
  \langle E\rangle_Z &= \langle H\rangle-
                        \langle H_\text{B}\rangle_\text{B}\\
                     &= \langle E\rangle + [\langle H_\text{SB}\rangle
                         +\langle H_\text{B}\rangle
                         -\langle H_\text{B}\rangle_\text{B}]
\end{aligned}
\end{equation}
where the index ``B'' denotes an average with respect to the bath
Hamiltonian $H_\text{B}$ only. This result differs from
the energy $\langle E\rangle=\langle H_\text{S}\rangle$ by the
term in the brackets which, generally, vanishes only in the absence of a
system-bath coupling. The coincidence between $\langle E\rangle$ and
$\langle E\rangle_Z$ for the harmonic oscillator subject to strictly
ohmic damping should therefore be considered as exceptional.

Nevertheless, we briefly sketch how one would obtain the specific heat from
the partition function because this will give us as a by-product an expression
for the entropy of the damped harmonic oscillator. For strictly ohmic damping
the product (\ref{ZHOe}) does not converge and in principle a high-frequency
cutoff for $\hat\gamma$ should be introduced. However, this divergence can
again be traced back to an infinite energy shift due to the environmental
coupling. We may shift the energy by an arbitrary amount $\Delta$ by
multiplying the partition function by $\exp(-\beta\Delta)$ without changing the
entropy or the specific heat. After performing an appropriate energy shift, we
arrive at an expression of the partition function valid even for strictly
ohmic damping \cite{lnp}, reading
\begin{equation}
  \bar Z = \frac{A}{\hbar\beta\omega_0}\left(\frac{2\pi}{\hbar\beta\omega_0}
  \right)^{\hbar\beta\gamma/2\pi}\Gamma\left(1+\lambda_+\right)
  \Gamma\left(1+\lambda_-\right) \,,
\end{equation}
where $\Gamma(z)$ is the gamma function, $\lambda_\pm$ are defined in
(\ref{lambda}), and $A$ is a constant whose precise
value is irrelevant for the following. By virtue of (\ref{HOS}) we
obtain for the entropy the result
\begin{equation}
  \label{SDHO}
  S=k_\text{B}\left[1-\ln(\hbar\beta\omega_0) +
  \frac{\hbar\beta\gamma}{2\pi}+
  g\left(\lambda_+\right) +
  g\left(\lambda_-\right)\right] \,,
\end{equation}
where we introduced the abbreviation
\begin{equation}
  g(z) = \ln[\Gamma(1+z)]-z\psi(1+z)
\end{equation}
with the digamma function $\psi(z)=\Gamma'(z)/\Gamma(z)$. In the absence of
damping, \ie $\gamma=0$, this reproduces the  result (\ref{HOSe}) for the
entropy of an uncoupled harmonic oscillator.

For very low temperatures, the entropy (\ref{SDHO}) vanishes like
\begin{equation}
  \label{Slt}
  S=\frac{\pi}{3}\frac{\gamma}{\omega_0}\frac{k_\text{B}^2T}{\hbar\omega_0}
  +O(T^3)
\end{equation}
as required by the Third Law of thermodynamics. By means of (\ref{CS}), the 
expression (\ref{CHO}) for the specific heat is recovered identically.

The low-temperature behavior (\ref{Slt}) is in agreement with the
expression derived by Ford and O'Connell on the basis of the free energy
\cite{ford05}. For the damped harmonic oscillator these authors 
found that the entropy vanishes in the limit of zero temperature also for 
more general forms of the bath density of states.

\section{Free quantum Brownian motion coupled to a heat bath: Is the
Third Law obeyed at zero temperature?}

According to the equipartition theorem the specific heat of a free
particle is $k_\text{B}/2$. In the limit of an infinite ``box'' this
represents even the correct quantum value, because the classical
and the quantum partition function become equal. In view of the fact that
the specific heat of an ideal gas thus remains non-zero down to the lowest
temperatures, one becomes curious to investigate the specific heat
of a free particle coupled to an environment when the coupling
strength is not assumed to vanish: Does quantum dissipation help to
restore the Third Law?

Free quantum Brownian motion has been addressed in
earlier work \cite{aslangul85,hakim85,gsi87,schramm87} wherein the main
focus centered on the role of free quantum diffusion
\cite{ingoldPR}. Interestingly enough, for ohmic dissipation the
quantum diffusion remains ``classical'', being proportional to time
$t$, except at zero temperature itself, where one finds a
logarithmic behavior in time $t$ \cite{aslangul85,hakim85,gsi87,schramm87}. At
finite temperatures this quantum behavior is observable at
intermediate times only \cite{aslangul85,jung}. One is thus tempted to 
conclude that
finite quantum dissipation will not be sufficient to cure the classical
behavior for the specific heat. Therefore, we shall next investigate
the behavior of the specific heat for free quantum Brownian motion in closer detail.

\begin{figure}
  \centerline{\includegraphics[width=0.8\textwidth]{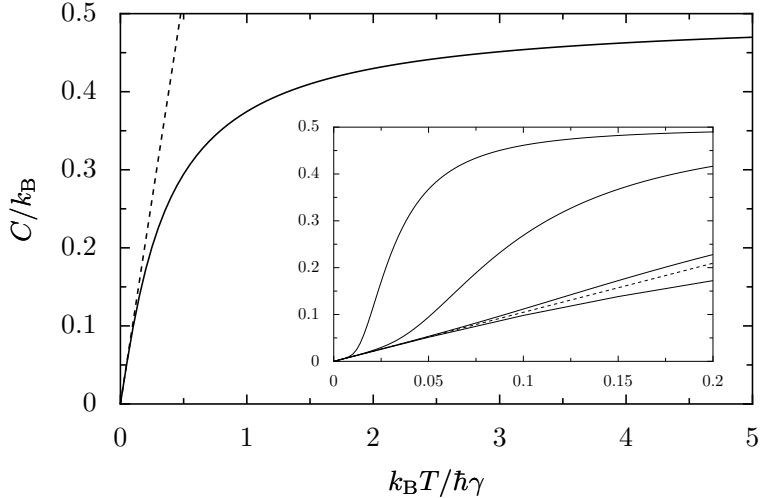}}
  \caption{The specific heat $C$ of a free quantum Brownian particle is shown as a
  function of the temperature $T$ for strictly ohmic friction of strength
  $\gamma$. The dashed line indicates the linear low-temperature behavior.
  In the inset, the modification due to a finite cutoff frequency
  $\omega_\text{D}$ is depicted for different cutoff scales, $\omega_\text{D}/\gamma=0.01,0.1,1,
  \text{and }\infty$ from the upper to the lower full line. The dashed line
  indicates again the linear low-temperature behavior.}
  \label{cfree}
\end{figure}

As already mentioned at the end of Section~\ref{pfe}, simply taking
the limit $\omega_0\to0$ in the results for the damped harmonic
oscillator is not without problems.  We therefore start our calculation
from the energy
\begin{equation}
  \langle E\rangle = \frac{\langle p^2\rangle}{2M}=
  \frac{1}{2\beta}\left[1+2\sum_{n=1}^\infty
  \frac{\nu_n\hat\gamma(\nu_n)}{\nu_n^2+\nu_n\hat\gamma(\nu_n)}\right]
\end{equation}
which can be obtained from the second moment of momentum $\langle
p^2\rangle$ by evaluating the integral (\ref{fnT}) by residues. Proceeding
as in Section~\ref{cdho}, one derives the specific heat of the free
particle in the presence of strictly ohmic damping
\begin{equation}
  \label{CFPo}
  \frac{C}{k_\text{B}}=\frac{1}{2}-\frac{\hbar\beta\gamma}{2\pi}
  +\left(\frac{\hbar\beta\gamma}{2\pi}\right)^2\psi'\left(1+
  \frac{\hbar\beta\gamma}{2\pi}\right)
\end{equation}
which for either $T\to\infty$ or $\gamma\to0$ yields $C=k_\text{B}/2$, as
expected from the equipartition theorem. On the other hand, for low
temperatures the specific heat tends to zero as
\begin{equation}
  \label{CFPlin}
  \frac{C}{k_\text{B}}=\frac{\pi}{3}\frac{k_\text{B}T}{\hbar\gamma}-
  \frac{4\pi^3}{15}\left(\frac{k_\text{B}T}{\hbar\gamma}\right)^3+O(T^5)
\end{equation}
in agreement with the Third Law of thermodynamics. The specific heat
(\ref{CFPo}) together with its linear low-temperature behavior are depicted
in the main part of Fig.~\ref{cfree} as full and dashed line, respectively.
Temperatures $k_\text{B}T\gg\hbar\gamma$ much larger than the damping
strength are required in order to restore the classical result
$C/k_\text{B}=1/2$.

As the damping strength already serves to set the temperature scale, it is
instructive to introduce a cutoff in the density of states (\ref{jom}) of the bath
oscillators in order to study how a reduction of the environmental
influence changes the dependence of the specific heat on temperature.
To this end, we introduce a high-frequency cutoff $\omega_\text{D}$ by
choosing the Drude model where
\begin{equation}
  \hat\gamma(z)=\frac{\gamma\omega_\text{D}}{z+\omega_\text{D}}\,.
\end{equation}
This environment leads to a specific heat
\begin{equation}
  \frac{C}{k_\text{B}}=\frac{1}{2}-\frac{\hbar\beta\gamma}{2\pi}
  \frac{1}{\sqrt{1-4\gamma/\omega_\text{D}}}\left[z_+\psi'(1+z_+)
  -z_-\psi'(1+z_-)\right] \,,
\end{equation}
where we have introduced the abbreviations
\begin{equation}
  z_\pm = \frac{\hbar\beta\omega_D}{4\pi}
  \left(1\pm\sqrt{1-\frac{4\gamma}{\omega_\text{D}}}\right)\,.
\end{equation}
Independently of the value of the cutoff frequency $\omega_\text{D}$, the
specific heat will go to zero linearly as stated in (\ref{CFPlin}).
However, for temperatures larger than the cutoff frequency, the suppression
of the specific heat due to the environmental coupling will be ineffective.
A reduction of the environment thus tends to restore the classical value
of $k_\text{B}/2$ for the specific heat. This effect is depicted in the inset
in Fig.~\ref{cfree}, where the cutoff frequency takes the values
$\omega_\text{D}/\gamma=0.01, 0.1, 1$ and $\infty$ from the upper to the
lower curve. The dashed curve represents the linear low-temperature
behavior which still dominates at temperatures below
$\hbar\omega_\text{D}/k_\text{B}$. In contrast to many phenomena in quantum
dissipation, here an increase of the coupling to the environment does not render the system
more classical. On the contrary, a stronger environmental coupling  makes the
dissipative quantum system behave more quantum mechanically and thus helps to
ensure the validity of the Third Law of thermodynamics,
\ie the vanishing of the specific heat with decreasing temperatures.

\section{Conclusions}
With this work we have explored the behavior of the specific heat for a
quantum system that is coupled to a heat bath with finite coupling
strength. Our findings are contrasted with the Third Law of
thermodynamics which generically predicts a vanishing of the
specific heat at low temperatures. For a harmonic oscillator the
presence of quantum dissipation  changes the well-known
Einstein-like behavior of an exponentially fast approach towards
zero specific heat into a power-law behavior with a
slope that increases with increasing coupling strength.
Even more intriguing is the behavior for a freely moving quantum
particle: While the quantum treatment in absence of dissipation
simply coincides with the classical behavior, \ie the specific
heat takes a constant value $C= k_B/2$, the role of finite quantum
dissipation is able to restore the Third Law, yielding a leading
linear temperature dependence. Quite counterintuitively,
its approach to the classical value occurs the faster the weaker
is the dissipation strength.

In contrast to common quantum statistical mechanics which
intrinsically is based on a vanishingly small coupling to the
environment, the finite coupling strength between the sub-system and
the bath causes some subtleties that must be recognized. As
made explicit in Section~\ref{evspfdho}, the thermodynamic
quantities depend on the procedure invoked in their definition:
the commonly used expression based on the partition function provides
results that generally do \textit{not} agree with the result obtained
from the corresponding quantum expectation value.  Interestingly
enough, in the strict ohmic limit (\ie in the absence of a high-frequency
cutoff) the specific heat for the damped harmonic oscillator
does not depend on the prescription employed. For the case of memory
friction, however, where a finite cutoff frequency is present, the
thermodynamic quantities depend both on the value of this cutoff
and the prescription used in their evaluation.

The results obtained for the low-temperature behavior of the specific heat of
simple quantum systems are not only of academic interest, but may turn out to
be relevant for experiments in nanoscience where one tests the quantum 
thermodynamics of small systems \cite{fqmt} that are coupled to an environment 
with a finite coupling strength.

\section*{Acknowledgment}

This work has been supported by  the Deutsche Forschungsgemeinschaft
(DFG)  (PH, SFB 486). Both authors like to congratulate Peter
Talkner for his first 60 years and his fine scientific career. We
both have heavily profited repeatedly and continue to strongly
profit from his insight and breadth of knowledge. May our present
ongoing  fruitful collaborations with him blossom further and even
intensify.


\begin{thebibliography}{99}
\bibitem{talknerFN}
The authors of reference \cite{talkner84} introduced two different
partition functions for a damped harmonic oscillator, see their
Eqs.~(4.19) and (5.11). The one which corresponds to our Eq.~(\ref{ZHOe})
is denoted as $Z'$ and is defined in their Eq.~(4.19).
\end{thebibliography}
\end{document}